\documentclass[preprint2]{aastex7}
\usepackage{CJK}

\begin{document}
\begin{CJK*}{UTF8}{gbsn}
\title{The Cosmic One-Eyed Smile: Revealing the Hidden Face of Mike Wazowski}

\author[0000-0001-6406-1003]{Chun Huang (黄淳)}\email{chun.h@wustl.edu}
\affiliation{Physics Department and McDonnell Center for the Space Sciences, Washington University in St. Louis; MO, 63130, USA;\\}
\correspondingauthor{Chun Huang}

\begin{abstract}
We present a novel and somewhat whimsical approach to pulsar hotspot modeling by drawing inspiration from the iconic one-eyed monster, Mike Wazowski, from \emph{Monsters, Inc.}. Utilizing X-ray high-quality timing data from NICER, we apply a Bayesian inference framework to model the X-ray pulse profile of PSR~J0437--4715. Our analysis employs a \emph{Wazowski Configuration} (WC) in which the conventional hotspot parametrization is replaced with a predefined image template, whose redness and size are adjusted to mimic temperature variations. The results reveal a configuration where two hotspots—one brighter and smaller in the north represents the energetic  ``University time Wazowski", and one larger yet cooler in the south represents the ``Monster, Inc. time Wazowski"—combine to produce the observed X-ray pulse profile. These findings not only demonstrate the sensitivity of pulse profile modeling to hotspot morphology but also open up the intriguing possibility that the X-ray emission of some pulsars may be interpreted as a cosmic homage to our favorite animated character.
\end{abstract}



\section{Introduction} 

\emph{Monsters, Inc.} and \emph{Monster University}, produced by the \texttt{PIXAR\textsuperscript{\textregistered}} studio, stand as landmarks in 3D animation and remain among the author's favorite films. The on-screen monsters---Sullivan, the lovable furry giant, and his best friend Mike Wazowski, the iconic one-eyed monster with an infectious smile---have long served as symbols of love and courage, inspiring many (myself included) to grow into, dare I say, good adults.

As both an astronomer and a theoretical physicist, the author felt an irresistible pull to search the cosmos for traces of our beloved monster friends. This paper primarily focuses on Mike Wazowski, who, in our view, might be hiding somewhere in the vast universe\footnote{Even though we have their official university website for reference: \url{http://michel.sales.online.fr/MU/monstersuniversity.com/edu/index.html}}. Although his signature presence has not yet been detected, recent advances in X-ray observations and cutting-edge instrumentation now allow us, for the first time, to propose a direct search for the hidden face of our old cinematic companion.

In this study, we explore one possible avenue to pinpoint his cosmic habitat. This approach takes us to the surface of another ``strong cosmic monster''---a neutron star, or technically, a rotation-powered millisecond pulsar (MSP)---where we search for its bright X-ray emissions from its surface. Past observations have struggled to fully reveal its features due to limitations in timing accuracy. However, thanks to the high-quality timing data from NASA's Neutron star Interior Composition Explorer (\emph{NICER}) \citep{NICER}, we can now apply the X-ray pulse profile modeling technique to perform detailed modeling of the mass, radius and the surface emission pattern of MSPs. This established technique deconstructs X-ray light curves and spectra to map temperature distributions across the star’s surface, allowing us to extract key parameters such as mass and radius~\citep[see, e.g.,][]{Morsink2007,B2013,2014ApJ...792...87P,Vincent2018}. NICER’s precision has led to several mass--radius measurements for pulsars such as PSR~J0030+0451~\citep{Miller2019,Riley2019,Vinciguerra2024} and PSR~J0740+6620~\citep{Miller2021,Riley_2021,Dittmann24,Salmi_2024}, thereby enhancing our understanding of neutron star structure and the surface emission of neutron star hotspots~\citep[see, e.g.,][]{Raaijmakers_2019,Raaijmakers2020,Raaijmakers2021,Huang2024,Huang:2024rvj,Rutherford2024}.

Detailed hotspot mapping has revealed that pulsar surfaces often exhibit complex temperature patterns. Observations of PSR J0030+0451, for example, uncovered distinct hotspot morphologies—one nearly circular and the other crescent-like—implying that the underlying magnetic field structure may extend beyond a simple centered dipole~\citep{Bilous_2019}. Theoretical investigations, including those by \citet{Chen_2020} and \citet{Kalapotharakos2021}, have explored how combinations of dipolar and quadrupolar magnetic fields can give rise to such intricate hotspot configurations. More recently, \citet{Vinciguerra2024} demonstrated that an off-center dipole, coupled with a nontrivial temperature distribution, could account for the observed X-ray features, thereby broadening the parameter space for modeling pulsar emissions.

In a playful yet rigorous twist on these conventional methods, we introduce a novel approach: replacing the conventional hotspot parametrizations with predefined image inspired by none other than Mike Wazowski. By deliberately adjusting the color temperature and spatial extent of these \emph{Wazowski Configuration} (WC) using bayesian inference technique, we explore whether such a whimsical configuration can still reproduce the key features of the X-ray observations. Although this approach may raise a smile, it is rooted in a serious attempt to probe the sensitivity of pulse profile modeling to hotspot morphology. 

In this work, we focus on the \emph{NICER} observation of PSR J0437--4715, the brightest millisecond pulsar known. This pulsar benefits from excellent radio observations and a reliable mass estimate obtained via Shapiro delay measurements \citep{Reardon_2024}. It is also a primary target of \emph{NICER} \citep{Choudhury_2024}, which has reported the presence of a ring-shaped hotspot alongside a circular hotspot on its surface. Interestingly, a recent study \citep{huang2025physicsmotivatedmodelspulsar} demonstrated that the pulsar's one-dimensional pulse profile can be alternatively explained by a hotspot configuration featuring an arch-shaped companion paired with a circular hotspot. This revelation sparked our curiosity, inspiring us to consider whether this observation might be interpreted---at least in part---as the cosmic imprint of our beloved character, with Mike Wazowski's unmistakable big smile and single eye mirroring the irregular hotspot.

This paper is organized as follows. In Section~\ref{sec:method}, we outline the theoretical framework and modeling techniques. Section~\ref{sec:X-ray} details our findings from the pulse profile modeling search of Mike Wazowski. Finally, Section~\ref{sec:conclusion} discusses the implications of our results, the limitations of our approach, and prospects for future research that unites rigorous astrophysical inquiry with a dash of interstellar whimsy.
\section{Method}
\label{sec:method}
In this section, we introduce the formalism behind the \emph{Wazowski Configuration} (WC) — a playful yet rigorous framework to model the hotspot features on pulsars. Figure~1 displays the standard Wazowski picture we are using, characterized by its distinctive green body, which serves as our baseline template.

\subsection{Defining the Redness Measure}

To quantitatively capture and adjust the redness in our \emph{Wazowski Configuration} image, so that the redness could quantitatively reflect the realistic temperature of the WC hot spot, we utilize Python's image processing libraries, specifically \texttt{Pillow} (PIL) and \texttt{NumPy}. This approach is analogous to using adjustment layers in Photoshop, where the red intensity can be controlled to simulate variations in temperature.

The process begins by loading the image with \texttt{Pillow} and converting it into a \texttt{NumPy} array, which facilitates the manipulation of individual color channels. We then separate the image into its red, green, and blue components, denoted by \(I_{\text{red}}\), \(I_{\text{green}}\), and \(I_{\text{blue}}\), respectively.

The \emph{redness measure}, \(\mathcal{R}\), is defined as:
\[
\mathcal{R} = \frac{I_{\text{red}}}{I_{\text{red}} + I_{\text{green}} + I_{\text{blue}}},
\]
where a higher value of \(\mathcal{R}\) indicates a region of higher temperature. Since the temperature is modeled using the image color of WC, we normalize the redness and then multiply it by a temperature normalization constant of \(10^7\) K on both WC hotspots. However, because we are fitting the 1D pulse profile, we argue that the relative temperatures of the two hotspots are more relevant for the modeling process.

To mimic the effect of enhancing the red intensity—as one would do in Photoshop—we adjust the red channel by multiplying it with a factor \(R\) corresponding to increased redness). After this multiplication, we clip the pixel values to ensure they remain within the valid 8-bit range (0--255). The enhanced red channel is then recombined with the original green and blue channels to produce the final image.

This method allows us to precisely control the redness of our WC hotspot images, simulating different temperature distributions across the pulsar's surface. The parameter \(R\) serves as our adjustable "rediness" control and is one of the key fitting parameters in our MCMC routine. This approach ensures a consistent and reproducible integration of the Mike Wazowski template into our modeling framework.

\subsection{Model Parameters and MCMC Fitting}

Our analysis models the pulse profile of PSR~J0437--4715 by fitting a modified hotspot pattern inspired by Mike Wazowski's iconic image. The fitting procedure employs a Markov Chain Monte Carlo (MCMC) approach using the \texttt{emcee} package. In our model, we characterize the hotspot configuration with ten key parameters. These parameters include the scale factor, which controls the overall size of the hotspot image, \(S_1\) and \(S_2\); the rotation degree \(A_1\) and \(A_2\), which governs each WC hotspot image orientation; the redness factors (\(R_1\) and \(R_2\)), which determine the local red intensity (and thus effective temperature) of the hotspots; and the center positions, specified by the spherical coordinates \((\theta_1,\phi_1)\) and \((\theta_2,\phi_2)\), that locate the hotspots on the stellar surface. Together with the observational inclination angle, \(\iota\), which determines the shape of the light curve, these parameters constitute the complete set of free parameters explored in this work.

Table~\ref{tab:params} summarizes the model parameters along with the uniform priors adopted for our MCMC fitting.

\begin{table*}
\centering
\caption{Model Parameters and Uniform Priors for the MCMC Fitting}
\label{tab:params}
\begin{tabular}{l l l}
\hline
\textbf{Parameter} & \textbf{Description} & \textbf{Prior Range} \\
\hline
\(\iota\) & Observer inclination angle & \(\mathcal{U}(0, \pi)\) \\
\(S_1\)           & Scale Factor controlling Hotspot 1 size       & \(\mathcal{U}(0.5, 2.0)\) \\
\(S_2\)           & Scale Factor controlling Hotspot 2 size       & \(\mathcal{U}(0.5, 2.0)\) \\
\(A_1\)         & Rotation degree that control the rotation of Hotspot 1  & \(\mathcal{U}(0, 2\pi)\) \\
\(A_2\)         & Rotation degree that control the rotation of Hotspot 1  & \(\mathcal{U}(0, 2\pi)\) \\
\(R_1\)         & Redness Factor for Hotspot 1 (temperature proxy)  & \(\mathcal{U}(0, 1)\) \\
\(R_2\)         & Redness Factor for Hotspot 2 (temperature proxy)  & \(\mathcal{U}(0, 1)\) \\
\(\theta_1\)    & Polar angle for Hotspot 1 center             & \(\mathcal{U}(0, \pi)\) \\
\(\phi_1\)      & Azimuthal angle for Hotspot 1 center           & \(\mathcal{U}(0, 2\pi)\) \\
\(\theta_2\)    & Polar angle for Hotspot 2 center             & \(\mathcal{U}(0, \pi)\) \\
\(\phi_2\)      & Azimuthal angle for Hotspot 2 center           & \(\mathcal{U}(0, 2\pi)\) \\
\hline
\end{tabular}
\end{table*}

These parameters allow us to flexibly adjust the simulated hotspot's size, shape, redness, and position, enabling a detailed exploration of how the Mike Wazowski-inspired features can reproduce the observed X-ray pulse profile.

\subsection{Light Curve Computation}

For our inference,  the X-ray light curves are computed from the resulting surface temperature profiles using the open-source \emph{X-PSI} package \citep{Riley2019}\footnote{\url{https://github.com/xpsi-group/xpsi.git}}. This package is based on a robust methodology developed over several studies \citep{Bogdanov_2019,Riley2019,Bogdanov_2021,Riley_2021} and typically represents hotspots as overlapping spherical caps with uniform temperatures. However, it also accommodates irregular hotspot shapes by allowing a detailed temperature distribution over the entire stellar surface~\citep[see, e.g.,][]{Das:2024jxc,huang2025physicsmotivatedmodelspulsar}. For clarity, the physical parameters used in our light curve computations are summarized in Table~\ref{tab1:xpsi_para}.

Due to the high computational cost associated with inference and limited computational resources, we coarse-grained the 300$\times$300 Temperature map to a 30$\times$30 representation of the WC configuration to reduce computational time. Consequently, our modeling is relatively crude. In standard modeling procedures, the background temperature outside the hotspot is typically treated as a free parameter; however, given our computational constraints, we fixed the background temperature to \(10^3\) K. This value roughly corresponds to the correct order of magnitude and reflects the contrast between the brightness of the hotspots and that of the background.
\begin{figure}
\centering
\includegraphics[width=0.3\textwidth]{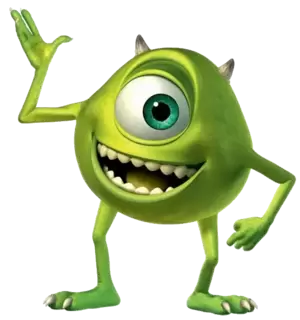}
\caption{The standard Mike Wazowski picture, featuring his iconic green body, used as the baseline template in our \emph{Wazowski Configuration} (WC) modeling.}
\label{fig:Mike}
\end{figure}

\begin{table}
\centering
\caption{Physical parameters for PSR~J0437--4715 used in the light curve computations.}
\label{tab1:xpsi_para}
\begin{tabular}{l c}
\hline
\textbf{Parameter} & \textbf{Value} \\
\hline
Mass, \(M\) & \(1.418\, M_{\odot}\) \\
Radius, \(R\) & 11.36 km \\
Distance, \(d\) & 156.98 pc \\
Spin Frequency, \(f\) & 174 Hz \\
Magnetic Field, \(B\) & \(5\times10^{8}\) G \\
Background Temperature T & $10^3$ K\\
\hline
\end{tabular}
\end{table}

\begin{figure*}
    \centering
    \includegraphics[width=\textwidth]{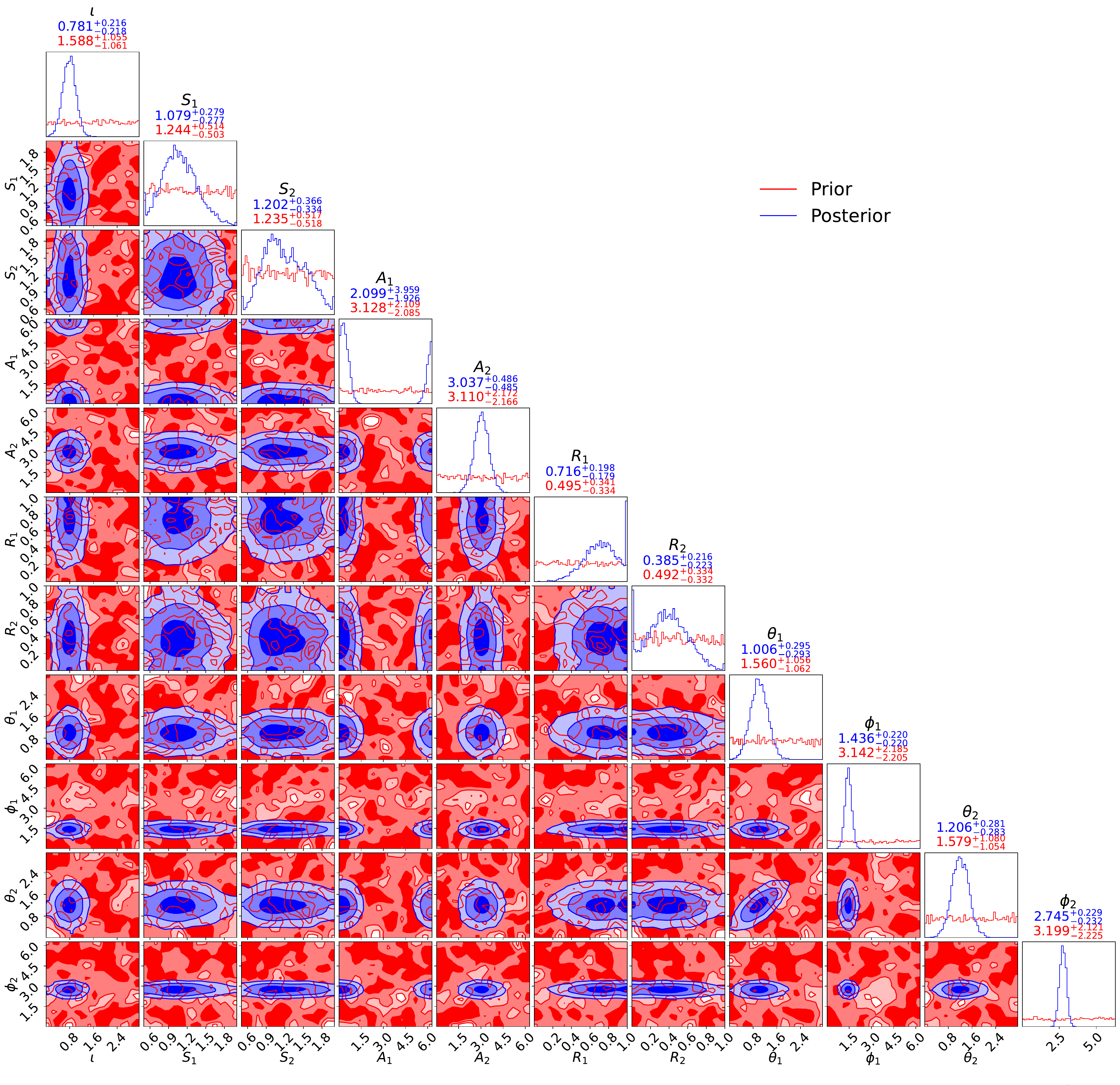}
    \caption{Corner plot comparing the prior (red) and posterior (blue) distributions for the model parameters used in the MCMC fitting of J0030's pulse profile. The parameters are: Observer inclination angle $\iota$, uniform prior in $[0,\, \pi]$, $S_1$ and $S_2$, the scale factors controlling hotspot sizes (uniform priors on $[0.5,\,2.0]$); $A_1$ and $A_2$, the rotation angles of the hotspots (uniform priors on $[0,\,2\pi]$); $R_1$ and $R_2$, the redness factors (uniform priors on $[0,\,1]$); and $(\theta_1,\phi_1)$ and $(\theta_2,\phi_2)$, the polar and azimuthal angles for the centers of Hotspot 1 and Hotspot 2 (with $\theta\in[0,\,\pi]$ and $\phi\in[0,\,2\pi]$). The contour levels, from deep to light, correspond to the 68\%, 84\%, and 98.9\% credible intervals, and the titles of the 1D marginal plots indicate the median and the range of the 68\% credible interval for each parameter.}
    \label{fig:posterior_corner}
\end{figure*}
\begin{figure*}
    \centering
    \includegraphics[width=0.8\textwidth]{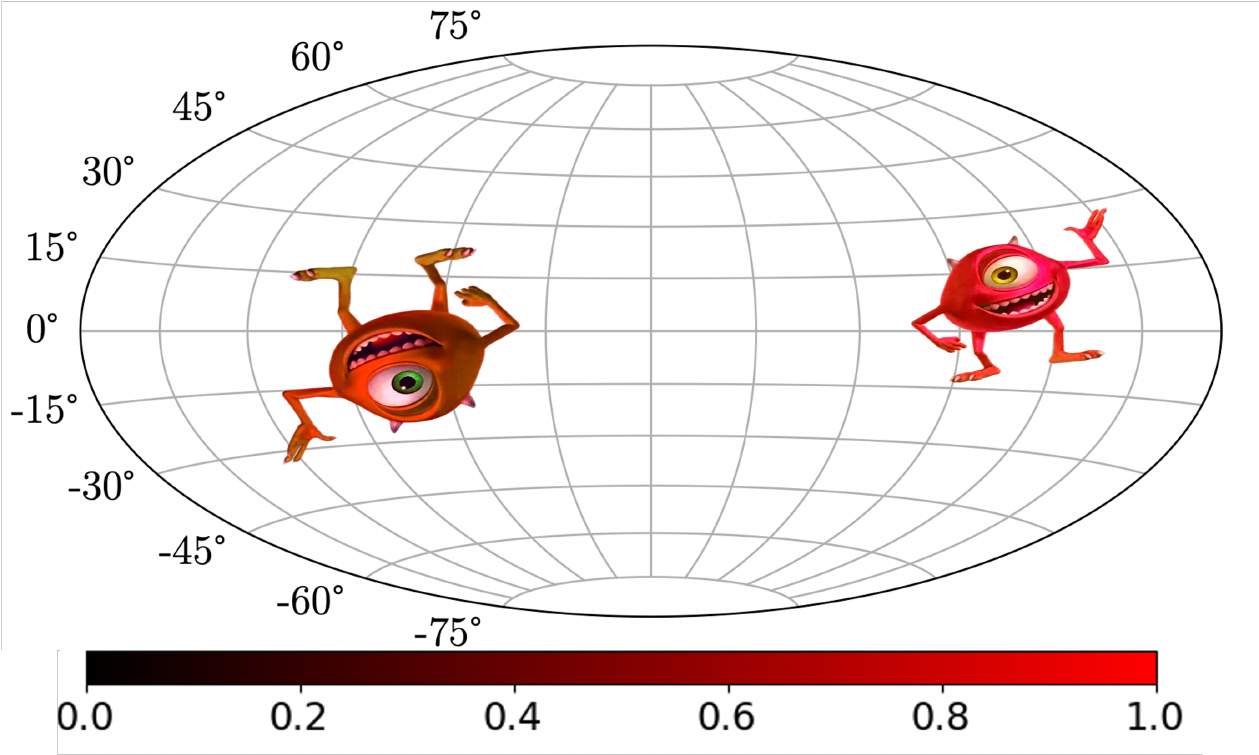}
    \caption{Hotspot configuration plot of the maximum likelihood sample that explained the J0437 observation, with two Mike Wazowski Configuration hotspots. The adjacent color bar (ranging from 0 to 1) represents the degree of rediness and hotness.}
    \label{fig:hotspot_mike}
\end{figure*}
\begin{table}
\centering
\caption{Maximum Likelihood Parameters for plotting the Hotspot Configuration in Figure \ref{fig:hotspot_mike}}
\label{tab:max_likelihood_params}
\begin{tabular}{l c}
\hline
\textbf{Parameter} & \textbf{Maximum Likelihood Value} \\
\hline
\(\iota\) & 0.81 \\
\(S_1\)     & 1.41 \\
\(S_2\)     & 0.98 \\
\(A_1\)     & 0.14 \\
\(A_2\)     & 2.98 \\
\(R_1\)     & 0.72 \\
\(R_2\)     & 0.35 \\
\(\theta_1\) & 0.99 \\
\(\phi_1\)   & 1.43 \\
\(\theta_2\) & 1.10 \\
\(\phi_2\)   & 2.78 \\
\hline
\end{tabular}
\end{table}

In summary, our approach leverages both MCMC fitting and advanced light-curve modeling to explore the unique hotspot configuration inspired by Mike Wazowski's ever-charming smile and single eye. This method refines standard techniques while adding a light-hearted twist to the rigorous field of pulsar astrophysics.

\section{Results: Pulse Profile Modeling Using the Wazowski Configuration}
\label{sec:X-ray}
In this section, we present the pulse profile modeling results using the WC hotspots by fitting the X-ray observation data of PSR J0437--4715. Figure~\ref{fig:posterior_corner} displays the posterior distributions of the modeling parameters obtained from our MCMC analysis. Compared to the original uniform priors, all parameters have decoupled and converged to well-defined posterior results.

The parameter \(A_1\) shows a multi-peak behavior. This is straightforward to understand: due to the periodic nature of the image rotation angle, any angle smaller than 0 wraps around to \(2\pi\), naturally producing multiple peaks. Additionally, the parameters \(A_1\) and \(A_2\) exhibit a clear \(\pi\) reversal. This implies that one of Mike Wazowski’s features—either his mouth or eye—flips from top to bottom when fitting the northern and southern hemispheres, a behavior that aligns well with the findings of \cite{huang2025physicsmotivatedmodelspulsar}.

Examining the ratio \(R_1/R_2\), we observe a value of approximately \(1/2\). In other words, one of the Mike Wazowski templates appears about twice as bright as the other, suggesting that one hotspot is more energetic. We suspect this could be the so-called ``university time shadow" of Mike Wazowski. Thus the southern naturally interpreted as the ``Monster Inc. time shadow" of our Mike Wazowski\footnote{One might argue that University Time Wazowski, being younger, is inherently more energetic.}.

A comparison of the \(\theta\) and \(\phi\) positions of the two hotspots reveals that they are closely aligned in the \(\theta\) (polar angle) direction, yet they are distinctly separated along the hemispherical divide—with one residing in the northern hemisphere and the other in the southern hemisphere. This spatial distribution is consistent with the results reported in \cite{huang2025physicsmotivatedmodelspulsar}.

Finally, while the positions of the hotspots are very well constrained by the X-ray observations, as evidenced by the tight posteriors in \(\theta\) and \(\phi\), the relative sizes (indicated by the posteriors of \(S_1\) and \(S_2\)) are less tightly constrained. This outcome demonstrates that our X-ray data can very precisely pinpoint the positions of the two Mike Wazowskis, even if it is more challenging to determine their exact sizes.

In Figure~\ref{fig:hotspot_mike}, we present the maximum likelihood configuration of the WC hotspots. As can be seen, both WC hotspots are positioned very close to the equatorial plane. Our analysis reveals that, consistent with the posterior distributions, one of the WC hotspots is significantly brighter than the other.
The WC hotspot in the southern hemisphere is larger in size, but its temperature is about half that of its northern counterpart. Conversely, the northern WC hotspot is much brighter while maintaining a smaller size. Notably, the two WC hotspots exhibit a rotation phase difference of \(\Delta A = A_1- A_2\approx \pi\), which effectively organizes the WC hotspot such that one primary region (the head of Mike Wazowski) is focused in the northern hemisphere and the other in the southern hemisphere.

Furthermore, the best-fitting observer inclination angle is approximately 0.81, in excellent agreement with radio observation inffered value for J0437 \citep{Reardon_2024}. The \(\phi\) angle difference between the two WC hotspots is measured to be 1.35, slightly lower than \(\pi/2\), indicating that they are not perfectly antipodal.

These findings suggest that the more energetic northern Wazowski, with its smaller and more luminous appearance, might be interpreted as the ``University Time Wazowski''---a playful nod to its vigorous emission. In contrast, the southern counterpart, larger in size and less energetic, represents the ``Monster, Inc. Time Wazowski.''

Overall, we successfully demonstrated that the X-ray pulse profile from PSR~J0437--4715 can be explained by two WC hotspots arranged in a specific configuration. This result opens up the intriguing possibility that the X-ray emission from this millisecond pulsar may, in part, be attributed to the presence of two Mike Wazowski-shaped hotspots. That hint the exsistence of this Monster inside of the universe.
\section{Conclusion and discussion}
\label{sec:conclusion}
In this work, we have explored a playful yet rigorous approach to pulsar hotspot modeling by introducing the \emph{Wazowski Configuration} (WC), inspired by the iconic one-eyed monster, Mike Wazowski. Our study shows that the X-ray pulse profile of PSR~J0437--4715 can be effectively reproduced by two hotspots arranged in a specific configuration: a brighter, compact hotspot in the northern hemisphere and a larger, cooler one in the southern hemisphere. The consistency between our modeling results and independent radio observations, along with the tight constraints on hotspot positions, lends credibility to our method.

The ability to explain the pulse profile with WC hotspots implies that the temperature distribution across the pulsar surface might be more irregular than predicted by conventional models of uniform spherical caps. In particular, the distinct brightness ratio, spatial separation, and phase differences between the two hotspots suggest that factors such as magnetospheric anisotropies or localized heating may play a significant role. More whimsically, the interpretation of the more energetic northern hotspot as the ``University Time Wazowski'' adds a playful dimension to our astrophysical reaserch, and connect the \emph{Monster, Inc.} and \emph{Monster University} Wazowski into a uniform strong space-time configuration --- an neutron star

Despite these intriguing findings, several limitations must be acknowledged. First, the WC framework employs a coarse-grained temperature map (reduced from a 300$\times$300 to a 30$\times$30 grid) to reduce computational demands, which may oversimplify the true complexity of the pulsar's surface temperature distribution. Second, fixing the background temperature at \(10^3\) K—while practical—limits the model's flexibility in capturing subtle variations in emission.

Future work could address these limitations by incorporating higher-resolution temperature maps and allowing the background temperature to be a free parameter. Expanding the parameter space to include more nuanced hotspot geometries, or even considering alternative templates inspired by other iconic figures, may reveal additional insights into the interplay between magnetic field topology and surface emission\footnote{Perhaps you can see it on this same day next year.}. Furthermore, another group is currently attempting to connect Mike Wazowski with gravitational lensing—a completely different approach \footnote{Means see you next year!}.

While our approach is grounded in established astrophysical techniques, its whimsical twist serves as a useful benchmark for our pulse profile modeling technique. Looking ahead, further investigations—perhaps incorporating multiwavelength analysis—may provide additional insights into the nature of these hotspots. In short, our work hints at the delightful possibility that the cosmos, too, may harbor its very own cosmic Mike Wazowski.
\begin{acknowledgments}
The author would like to extend heartfelt thanks to \texttt{PIXAR\textsuperscript{\textregistered}.} Studio, whose imaginative creations not only made my childhood magical but also inspired the unforgettable characters that continue to spark my creativity\footnote{There are many characters to which I am equally grateful. From the first \emph{Toy Story} to recent \emph{Inside out 2}.}. The author acknowledges the computational resources provided by the WashU Physics Cluster and the support from NASA grant 80NSSC24K1095.
\end{acknowledgments}

%



\bibliography{sample7}{}
\bibliographystyle{aasjournalv7}


\end{CJK*}
\end{document}